\documentclass[10pt]{article}
\usepackage[affil-it]{authblk} 
\usepackage[font=small,floatrowsep=qquad,captionskip=5pt]{floatrow}
\usepackage{booktabs}
\DeclareCaptionFont{mysize}{\fontsize{10}{12}\selectfont}
\usepackage{sectsty}
\sectionfont{\fontsize{10}{12}\selectfont}
\subsectionfont{\fontsize{10}{12}\selectfont}
\subsubsectionfont{\fontsize{10}{12}\selectfont}
\usepackage{epsfig, keyval, chicago}
\usepackage[font=large,labelfont=bf,font=bf]{caption}
\usepackage[latin1]{inputenc}
\usepackage{amsmath, amssymb}
\DeclareFontFamily{U}{rcjhbltx}{}
\DeclareFontShape{U}{rcjhbltx}{m}{n}{<->rcjhbltx}{}
\DeclareSymbolFont{hebrewletters}{U}{rcjhbltx}{m}{n}
\let\aleph\relax\let\beth\relax
\let\gimel\relax\let\daleth\relax
\DeclareMathSymbol{\aleph}{\mathord}{hebrewletters}{39}
\DeclareMathSymbol{\beth}{\mathord}{hebrewletters}{98}
\DeclareMathSymbol{\gimel}{\mathord}{hebrewletters}{103}
\DeclareMathSymbol{\daleth}{\mathord}{hebrewletters}{100}
\DeclareMathSymbol{\lamed}{\mathord}{hebrewletters}{108}
\DeclareMathSymbol{\mem}{\mathord}{hebrewletters}{109}
\DeclareMathSymbol{\ayin}{\mathord}{hebrewletters}{96}
\DeclareMathSymbol{\tsadi}{\mathord}{hebrewletters}{118}
\DeclareMathSymbol{\qof}{\mathord}{hebrewletters}{114}
\DeclareMathSymbol{\shin}{\mathord}{hebrewletters}{152}
\usepackage{latexsym, rotating, hyperref}
\usepackage{grffile}
\usepackage{verbatim}
\usepackage{bm}
\usepackage{parskip}
\usepackage{wrapfig}
\usepackage{rotating}
\usepackage{pdfpages} 
\usepackage{graphicx}
\usepackage{geometry}
\usepackage{longtable}
\usepackage{lscape}
\begin{document}
\makeatletter 
\renewcommand\@biblabel[1]{#1} 
\makeatother %
\makeatletter
\newcommand\footnoteref[1]{\protected@xdef\@thefnmark{\ref{#1}}\@footnotemark}
\makeatother

\newcommand{\md}{{\Phi}}
\newcommand{\mdna}{{\Phi}\na}
\newcommand{\emdna}{\widetilde{\Phi}\na}
\hfuzz5pt 
\vfuzz5pt 

\vspace*{0.35in} 

\begin{flushleft}
{\Large
\textbf\newline{Global estimation of unintended pregnancy and abortion  using a Bayesian hierarchical random walk model.}
}
\newline
\bigskip \\
Jonathan Bearak\textsuperscript{1,*}, Anna Popinchalk\textsuperscript{1}, Bela Ganatra\textsuperscript{2}, Ann-Beth Moller\textsuperscript{2}, \"{O}zge Tun\c{c}alp\textsuperscript{2}, Cynthia Beavin\textsuperscript{1}, Lorraine Kwok\textsuperscript{1}  and Leontine Alkema\textsuperscript{3}

\bigskip 
1 Guttmacher Institute, New York, NY, USA
\\
2 Department of Sexual and Reproductive Health and Research, including UNDP/UNFPA/UNICEF/WHO/World Bank Special Programme of Research, Development and Research Training in Human Reproduction (HRP), World Health Organization, Geneva, Switzerland
\\
3  Department of Biostatistics and Epidemiology, University of Massachusetts Amherst, Amherst, MA, USA (L Alkema PhD)
\\
\bigskip
* Contact: jbearak@guttmacher.org

\bigskip

\end{flushleft}

\section*{Abstract}
Unintended pregnancy and abortion estimates are needed to inform and motivate investment in global health programmes and policies. Variability in
the availability and reliability of data poses challenges for producing estimates. We developed a Bayesian model that simultaneously estimates incidence of unintended pregnancy and abortion for 195 countries and territories. Our modelling strategy was informed by the proximate determinants of fertility with (i) incidence of
unintended pregnancy defined by the number of women (grouped by marital and contraceptive use status) and their respective pregnancy rates, and (ii) abortion incidence defined by group-specific pregnancies and propensities to have
an abortion. Hierarchical random walk models are used to estimate country-group-period-specific pregnancy rates and propensities to abort.

\section*{Funding}
UK Aid from the UK Government; Dutch Ministry of Foreign Affairs, \\ 
UNDP/UNFPA/UNICEF/WHO/World Bank Special Programme of Research, Development and Research Training in Human Reproduction (HRP); The Bill and Melinda Gates Foundation. 

\copyright 2020 The Authors. This is an Open Access article published under the CC BY NC ND 3.0 IGO license which permits users to download and share the article for non-commercial purposes, so long as the article is reproduced in the whole without changes, and provided the original source is properly cited. This article shall not be used or reproduced in association with the promotion of commercial products, services or any entity. In any use of this article, there should be no suggestion that WHO or Guttmacher endorses any specific organisation, products or services. The use of the WHO logo is not permitted. The authors alone are responsible for the views expressed in this article and they do not necessarily represent the views, decisions or policies of the institutions with which they are affiliated. This notice should be preserved along with the article's original URL.

		\newpage
	\tableofcontents
	
	\newpage

	
	\newcommand{\women}{W} 
	\newcommand{\births}{B} 
	
	\newcommand{\logN}{\mbox{logN}}
	\newcommand{\minrate}{0.0001}
	\newcommand{\maxrate}{5}
	\newcommand{\logminrate}{\log\minrate}
	\newcommand{\logmaxrate}{\log\maxrate}
	
	\newcommand{\npreg}{\Omega}
	\newcommand{\pregrate}{\omega}
	\newcommand{\nabo}{\Psi}
	\newcommand{\aboprop}{\alpha}
	\newcommand{\birthrate}{\theta}
	\newcommand{\nbirths}{\Theta}
	
	\newcommand{\totalvar}{\sigma^2}
	\newcommand{\totalsd}{\sigma} 
	\newcommand{\abosd}{\sigma}
	\newcommand{\upsd}{\eta}
	\newcommand{\upsdchange}{\upsilon}
	\newcommand{\upsdmin}{0.01}
	\newcommand{\upsdmax}{3}
	\newcommand{\subsetofcountries}{\mathcal{R}}
	\newcommand{\countriesinregion}{c \in \subsetofcountries}
	\newcommand{\region}{^{(region)}} 
	\newcommand{\world}{^{(world)}}
	\newcommand{\nonrep}{^{nonrep}}


	\newcommand{\errormult}{\lambda}
	\newcommand{\propabo}{\Upsilon}
	\newcommand{\propup}{\Lambda}
	\newcommand{\up}{\Psi}
	\newcommand{\propub}{\Theta}
	\newcommand{\ub}{\Omega}
	\newcommand{\superregion}{^{(\text{super-region})}}

\section{Introduction}
This manuscript provides the detailed technical description of the modeling approach used by Bearak et al. (forthcoming)\footnote{Bearak J, Popinchalk A,  Ganatra B, Moller AB, Tunçalp O, Beavin C, Kwok L and Alkema L (forthcoming). Unintended pregnancy and abortion by income, region, and the legal status of abortion: estimates from a comprehensive
model for 1990-2019. The Lancet Global Health} to estimate global incidence of unintended pregnancies and abortions. The paper is organized as follows: section~\ref{sec-data} introduces the data available for estimation. The model set-up is described next. We conclude with information regarding model validation. 

\section{Data}\label{sec-data}
\subsection{Intention Data}
Intention data were compiled from country-based surveys, or from one-time studies found through a literature search. We obtained data on the share of pregnancies and births unintended for 139 countries from the Demographic and Health Surveys (DHS), \footnote{The DHS Program. Demographic and Health Surveys. USAID http://dhsprogram.com/publications/citing-dhs-publications.cfm.} the Multiple Indicator Cluster Surveys (MICS), \footnote{Multiple Indicator Cluster Survey. UNICEF http://mics.unicef.org/surveys} the Reproductive Health Surveys (RHS), \footnote{Reproductive Health Surveys. Centers for Disease Control and Prevention, 2019} and from published studies. For Mexico and the United States, we obtained additional data from the National Survey of Demographic Dynamics (ENADID), \footnote{Encuesta Nacional de la Dinamica Demografica (ENADID). Instituto Nacional de Estadistica y Geografia (INEGI), 2018} and from the National Survey of Family Growth (NSFG), \footnote{National Survey of Family Growth. Centers for Disease Control and Prevention (CDC), National Center for Health Statistics (NCHS), 2018} respectively. For published studies, we searched PubMed and Google Scholar from May 2018 to May 2019 for available literature for each country in our analysis for the time period 1990-2018, with the search terms ``unplanned birth,'' ``unintended pregnancy,'' ``unwanted pregnancy,'' ``unplanned pregnancy,'' ``pregnancy intention,'' ``unintended births,'' and ``unwanted birth.''

\subsection{Abortion Data}
We obtained abortion data from published studies and from official statistics. We obtained official statistics from ministries of health and national statistical offices, or, if not otherwise available, from the United Nations Statistical Division (UNSD) Demographic Yearbook.\footnote{United Nations Statistical Division. UNSD -- Demographic and Social Statistics.} Official statistics on abortion may be incomplete due to various issues including the legality of abortion, differing reporting requirements across countries, and underreporting.\footnote{Sedgh G, Henshaw S. Measuring the Incidence of Abortion in Countries With Liberal Laws. In: Methodologies for Estimating Abortion Incidence and Abortion-Related Morbidity: A Review. New York: Guttmacher Institute, 2010: 23-30.} For example, abortion statistics may exclude procedures that occur outside the public sector. Where abortion is restricted, studies show that substantial numbers of illegal abortions take place.\footnote{Sedgh G, Sylla AH, Philbin J, Keogh S, Ndiaye S. Estimates of the Incidence of Induced Abortion And Consequences of Unsafe Abortion in Senegal. Int Perspect Sex Reprod Health 2015; 41: 11.} \footnote{Chae S, Kayembe PK, Philbin J, Mabika C, Bankole A. The incidence of induced abortion in Kinshasa, Democratic Republic of Congo, 2016. PLoS ONE 2017; 12: e0184389.}\footnote{Polis CB, Mhango C, Philbin J, Chimwaza W, Chipeta E, Msusa A. Incidence of induced abortion in Malawi, 2015. PLoS ONE 2017; 12: e0173639}\footnote{Sully EA, Madziyire MG, Riley T, et al. Abortion in Zimbabwe: A national study of the incidence of induced abortion, unintended pregnancy and post-abortion care in 2016. PLOS ONE 2018; 13: e0205239.} In order to assess whether official statistics included all abortions, data from official statistics were determined to be complete through a data classification process outlined in the previously published protocol.\footnote{Bearak J, Popinchalk A, Sedgh G, Ganatra B, Moller AB, Tun\c{c}alp \"{O} \& Alkema L. Pregnancies, abortions, and pregnancy intentions: a protocol for modeling and reporting global, regional and country estimates. Reproductive Health 2019} Of 104 countries with abortion data, 65 had a complete datum in one or more years. For published studies, we searched PubMed and Google Scholar from January 2018 to May 2019 for studies for each country in our analysis for the time period 1990-2018. Our search terms included ``abortion incidence,'' ``abortion estimates,'' ``termination of pregnancy,'' ``induced abortion,'' and ``menstrual regulation,'' followed by, one by one, the name of each country.

\subsection{Data Classification}
A data classification process (detailed in previously published study protocol) was applied to all available data. This process was designed to increase transparency in the treatment of all available information, and to ensure consistency in how the the model-based estimates incorporated information on data sparsity and quality.

For the data on pregnancy and birth intention, the classification process reviewed the study population, sample, unit of analysis, measure of pregnancy intention and whether or not we had access to the microdata for all available data. This classification allowed us to incorporate additional error terms or bias terms, or treat the data differently as needed. In brief: Information on the percent of births unintended were treated as point estimates, whereas information on the percent of pregnancies unintended were treated as minimum estimates due to abortion under-reporting. Where data pertained only to married women, the data were used to inform the percent unintended among married women, as well as the minimum percent unintended to all women. For published reports from DHS or RHS surveys which reported on the percent of births unintended in the last three of five years, to allow for the possibility of temporal response bias, we computed bias terms from DHS micro-data for comparability with the year-specific estimates we computed from the microdata. We also extracted information on the percent of births unintended separately by marital status from DHS and MICS surveys; since dates of union dissolution and were not recorded in those surveys, we constructed ranges (in contrast to point estimates) by calculating the percent of women whose marital status at the time of birth was not ascertainable. Finally, for data that used the London Measure of Pregnancy Intention, since it is a measure that uses a 12-item scoring system to categorise pregnancies according to whether the woman said she planned the pregnancy, did not plan it, or was ambivalent about it, we collapsed responses into two categories of ``unwanted,'' and ``planned.'' We then computed ranges since these two categories are extreme relative to the (un)intended dichotomy. More detail on this classification of the pregnancy and birth intention data can be found in Figure 3 (page 7) in the published protocol.\footnote{Bearak JM, Popinchalk A, Sedgh G, et al. Pregnancies, abortions, and pregnancy intentions: a protocol for modeling and reporting global, regional and country estimates. Reprod Health 2019; 16: 36.}

For abortion data, the reliability of the data can vary widely so each datum was classified to determine how it informed the estimates in the statistical model. Similar to the data on pregnancy and birth intention, the classification process reviewed the estimation method, sample and source of the data. We developed distinct classification processes for data which came from published studies and data which came from official statistics.

For abortion data from published studies the sample and methodology were reviewed, and consequently error terms or bias terms were incorporated. Specifically, for published studies using indirect methods, we input the study's estimated number of abortions from the direct component (e.g., the number of complications from illegal abortions to be treated in hospitals) and the indirect component (e.g., the percent of all abortions that this represented) separately, to allow for larger error in the indirect component. As with previous abortion incidence estimates, data from surveys of women were treated as minimum-only estimates, except for Central Europe, Eastern Europe, Central Asia and the Caucasus, where sufficient information comparing estimates based on surveys of women to reliable official statistics were available. As per the protocol, we computed a bias term using a multi-level model: this indicated that approximately two-thirds of abortions were reported in surveys of women in these regions, the same figure used in previous approaches.\footnote{Sedgh G, Bearak J, Singh S, Bankole A, Popinchalk A, Ganatra B, Rossier C, Gerdts C, Tun\c{c}alp \"{O}, Johnson BR, Johnston HB \& Alkema L. 2016. ``Induced abortion 1990 to 2014: Global, regional, and subregional trends.'' \textit{The Lancet.}}

Due to issues around abortion under-reporting, official abortion statistics have historically been assessed for completeness. Thus, data from official statistics were reviewed through a series of six hierarchical questions to ascertain whether the data should be treated as a point estimate or minima. Minimum estimates inform the model that the true abortion rate is no less than the observed rate. In addition, where it was possible that the statistic included spontaneous abortions, we computed a range (if the official statistics were otherwise classified as complete) or a minimum (if the official statistics were otherwise classified as incomplete) with a lower bound based on .1 miscarriages for every induced abortion and .2 miscarriages for every live birth. More detail on the classification of abortion data can be found in Table 1 (page 6) and Figure 2 (page 5) in the published protocol, however we summarize the six hierarchical questions below, with information on each country which has one or more years of official statistics classified as a minima provided in Table~\ref{tb:minima}.

\textit{Legal abortion is not broadly available:} One of the criteria for official statistics to be complete is assessing whether abortion is broadly available in a country. If not, official statistics are treated as incomplete. In 19 countries, abortion was prohibited altogether or available only to protect health, and thus one or more years were treated as minima (Table~\ref{tb:minima}). In 16 countries, this applied to all official statistics  (Bangladesh, Bhutan, Botswana, Burundi, Chile, Costa Rica, Dominican Republic, Hong Kong, Mexico, Myanmar, Panama, Peru, Poland, Togo, Tonga, Zimbabwe). In 6 of these countries, complete data were available from an in-country study (Bangladesh, Chile, Dominican Republic, Mexico, Peru, Zimbabwe). In countries where the legal status of abortion changed during the analysis period, if the law became more liberal, data from the years abortion was restricted were treated as minima (Table~\ref{tb:minima}2: Portugal through 2007, Spain through 2009, Switzerland through 2001). 

\textit{Government acknowledges that its statistics are incomplete:} Among countries with incomplete official statistics, 24 countries acknowledged that their statistics were incomplete for one or more years, and for 21 countries, this pertained to all years. This information was gathered through country reports, data collection questionnaires, or during the country consultation process (Albania, Armenia, Australia, Azerbaijan, Bhutan, Bosnia and Herzegovina, Canada [1999-2010], Croatia, Cuba [2002-2017], Germany [1990-1995], Kyrgystan, Macedonia, Mexico, Moldova, Nepal, Netherlands [2010], Panama, Puerto Rico, and Serbia, South Africa, Tunisia, and the United States). Of the 21 countries for which all years of official statistics were acknowledged to be incomplete, other sources of abortion counts were available for nine of these countries.\footnote{Specifically, in the United States, complete statistics are obtained from a census conducted by the Guttmacher Institute; in Australia, Mexico and Nepal, indirect estimates were available from the scientific literature; and finally, surveys of women were available (incorporated into the model with bias terms) were available from Armenia, Azerbaijan, Kyrgystan, Moldova, Serbia.}

\textit{Official statistics are below an estimate from a survey of women: } As stated in the protocol, when a country has both a survey of women and official statistics, if the survey of women indicates a higher abortion rate than counted in the official statistics, all official statistics are treated as minima. In countries within Central Asia, the Caucasus, and, among European countries, the post-Soviet states, women's self-reports were collected as part of the Reproductive Health Surveys (RHS) in several countries. In 12 countries, we treated data as incomplete because a higher estimate was reported in a survey of women (Albania, Armenia, Georgia, Kazakhstan, Moldova, Romania, Russia, Taiwan, Tajikistan, Turkmenistan, Ukraine, Uzbekistan; this also applies to Azerbaijan and Kyrgystan, which were classified as incomplete as per the preceding rule.)

\textit{A sizeable portion of abortions occur outside the formal health sector:} In 17 countries, evidence indicated that a sizeable portion of abortions occurred outside the formal health sector (Azerbaijan, Croatia, Greece, Hong Kong, India, Israel, Mexico, Moldova, Myanmar, Panama, Nepal, Peru, Poland, Romania, Serbia, South Africa, Vietnam [1996-2016]), seven of which were classified as minima already per reasons above, for an additional 10 countries. Of these additional 10, in one case (India), a complete count was available from an in-country study. This information was gathered through data collection, published studies, and official reports. 

\textit{Implausible levels or trends imply a country's data are incomplete: } For two other countries, external data or sources indicate that official statistics are incomplete due, consistent with implausible levels or trends. This affects all official statistics in Italy, and official statistics after 1991 in China (see Appendix Table~\ref{tb:minima} for additional details).

\subsection{Additional notes on abortion data classification}

In 22 of the countries which met any of the above criteria, spontaneous and induced abortions were reported together. In those instances, minima were adjusted downwards to take into account the spontaneous abortions. Specifically, one-tenth of the abortions were subtracted, and then the number was further reduced by a number equal to one-fifth of live births, and this lower number was used as a minimum number of induced abortions.

Finally, during the data classification process, we made a minor addendum to the protocol on how we classify subnational studies on abortion incidence. The protocol states that additional error is modeled for subnational surveys (Bearak et al 2019). However, data from nationally representative studies in Sub-Saharan Africa, with abortion rates for major cities in addition to national rates, indicates that abortion rates in major cities are higher than  national rates. We therefore revised the protocol to prevent over-estimation based on subnational studies. Specifically, we have classified data from any subnational studies of major cities as maxima. This decision affects two observations: a  study for the Democratic Republic of the Congo -- Kinshasa and a subnational study from Burkina Faso. In additional, we excluded women's reports from a 1996 RHS with subnational estimates in Russia, which suggests an increase compared to the previous nationally representative datum, whereas the nationally represents surveys, as well as the official statistics, show declines throughout the analysis period.

\subsection{Data Availability}

In total, we obtained 2,413 data points on unintended fertility and/or abortion from 166 countries for the time period 1990-2019. We compiled 516 observations on unintended fertility for 139 countries: 400 from survey micro-data, 51 from RHS or DHS reports where we did not have access to the micro-data, and 65 from published studies. For abortion data, we obtained 1,898 observations from 105 countries: 785 from official statistics classified as complete for 40 countries, 48 from published studies for 25 countries, 27 from surveys in 11 countries, and an additional 1,013 observations from 79 countries where the data were treated as minima or maxima.

As mentioned above, because some countries had one kind of data but not the other, we had abortion or intention data for 166 countries and territories. Excluding data treated as minima/maxima, the corresponding figures are 1,293 observations from 153 countries. We had both intention and abortion data for 52 countries; excluding countries with minima/max-only data, the number is 52.

For data availability by region, in almost all SDG regions, at least 95\% of all women of reproductive age were represented by all available data. The one exception was Oceania, excluding Australia and New Zealand, where coverage was only 89\%. New Zealand Australia have both unintended fertility and abortion data, and thus appear to have the most coverage, however there are only two countries in this region. Among other regions, Central and Southern Asia had the highest coverage for both unintended fertility and abortion data, at 93\% of women of reproductive age. 

In Central and Southern Asia, 93\% of countries in the region had data on unintended fertility. Sub-Saharan Africa had the highest number of data points on unintended fertility, although this represented a slightly smaller proportion of countries in the region, 84\%. Oceania, excluding New Zealand and Australia, and North America and Europe, had the lowest percent of countries with data on unintended fertility at 50\% and 51\%, respectively.   

The region with the most observations for all available data was North America and Europe with 1,018 data points. Due to the high number of countries in that region that report official abortion statistics, 96\% of those observations were abortion data, representing 90\% of countries in the region. About a third of those were treated as minima. Oceania, excluding Australia and New Zealand, and Western Asia and North Africa had the lowest number of abortion data, with only 12\% and 28\% of countries in these regions having abortion data. 

\begin{table}
    \includegraphics[page=1,scale=0.4,trim=5cm 14cm 5cm 2cm]{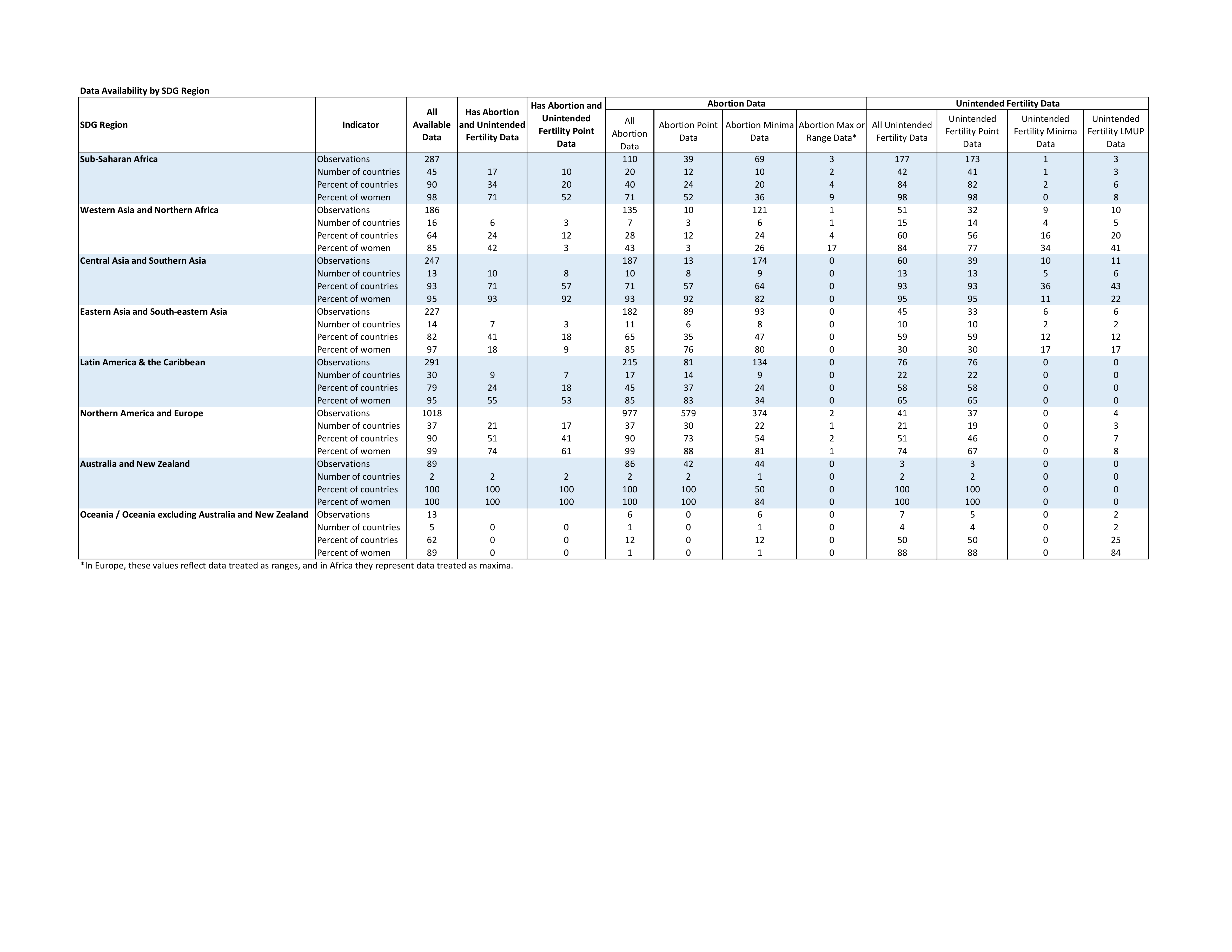}
    \caption{Data availability by SDG region.}
\end{table}

\begin{table}
    \includegraphics[page=1,scale=0.4,trim=5cm 14cm 5cm 2cm]{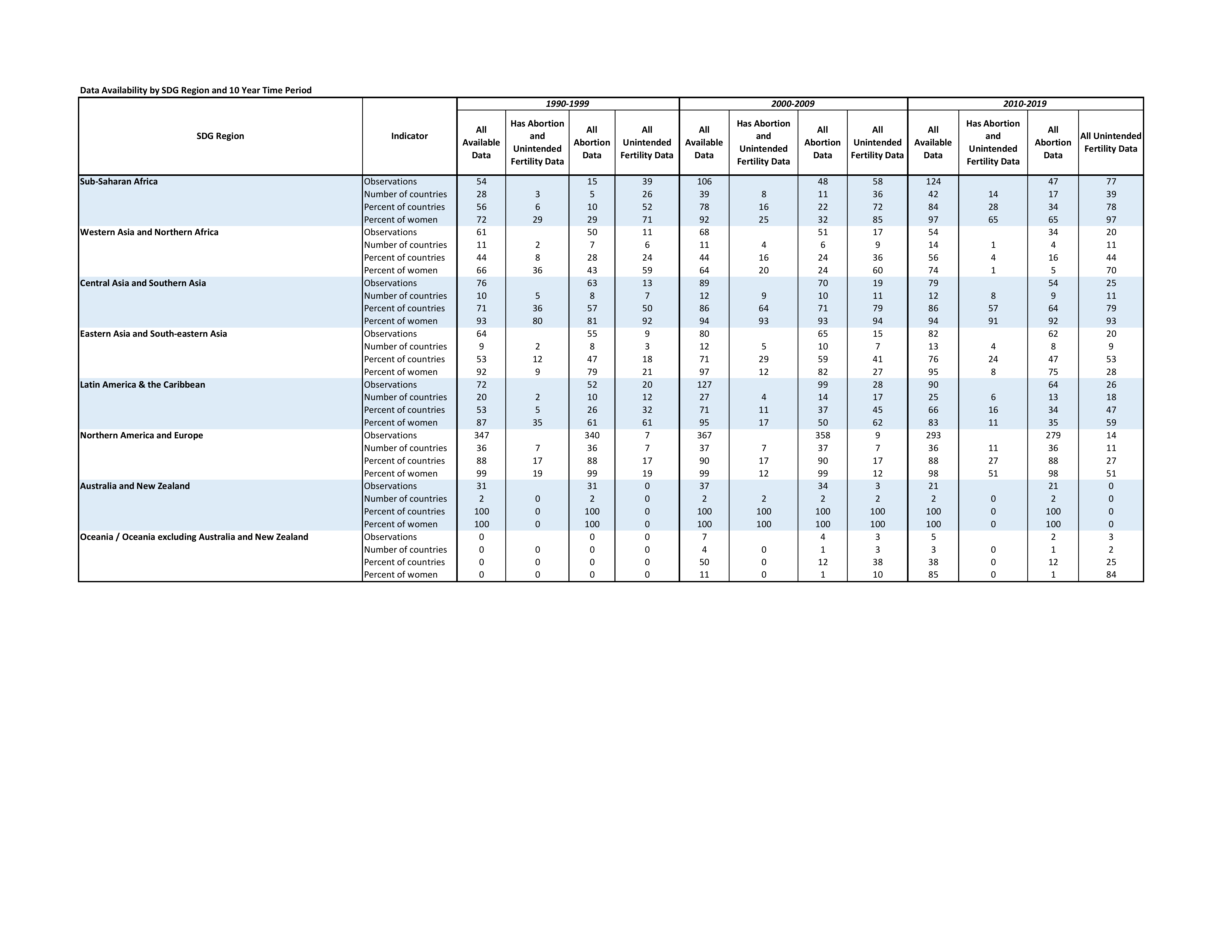}
    \caption{Data availability by SDG region and 10-year time period.}
\end{table}

\subsection{Marital Status Data}

Where possible we gathered data on unintended fertility by marital status, primarily with DHS and MICS survey data. For surveys of all women, we calculated the percent of marital births that were intended if a marital history was also available in the microdata. For surveys of ever married women, data on unintended fertility informed the percent of marital births intended. 

We utilized abortion by marital status data for official statistics or surveys with this information. Due to varying definitions and practices around marriage, in union, and/or cohabitation, data were treated as minima, point estimates, or maxima, depending on the categories of marital status presented, versus the definition of marital status in the corresponding countries. 

\section{Model}

\subsection{Notation}
In the model description, Greek letters refer to unknown parameters, whereas Roman letters refer to variables that are known or fixed, including data (lowercase) and estimates provided by other sources or the literature (uppercase).  

Countries are indexed by $c = 1,\hdots, C$ with C = 195 and calendar years by $t =1, \hdots, T$ referring to years 1990 to 2019. Five year periods are indexed by $p = 1, \hdots, P$ referring to each five-year period between 1990-2019. The five-period period for year $t$ is denoted by $p[t]$. 	For period $p$ before or after $p_0$, we define neighboring period $p^*[p] = p-1$ for $p>p_0$ and $p^*[p] = p+1$ for $p<p_0$.

	Truncated normal and $t$ distributions are used in the model specification, denoted as follows: if $a \sim t_{3}(b,c)T(d,e)$, then $\log(a)$ follows a t distribution with mean $\log(b)$, variance $c$, degrees of freedom 3, and $a$ is constrained to values in between $d$ and $e$. 

\subsection{Groupings of women by marital status and family planning}
\begin{figure}[htbp]
    \centering
    \includegraphics[scale=0.20,trim=9cm 0cm 0cm 0cm]{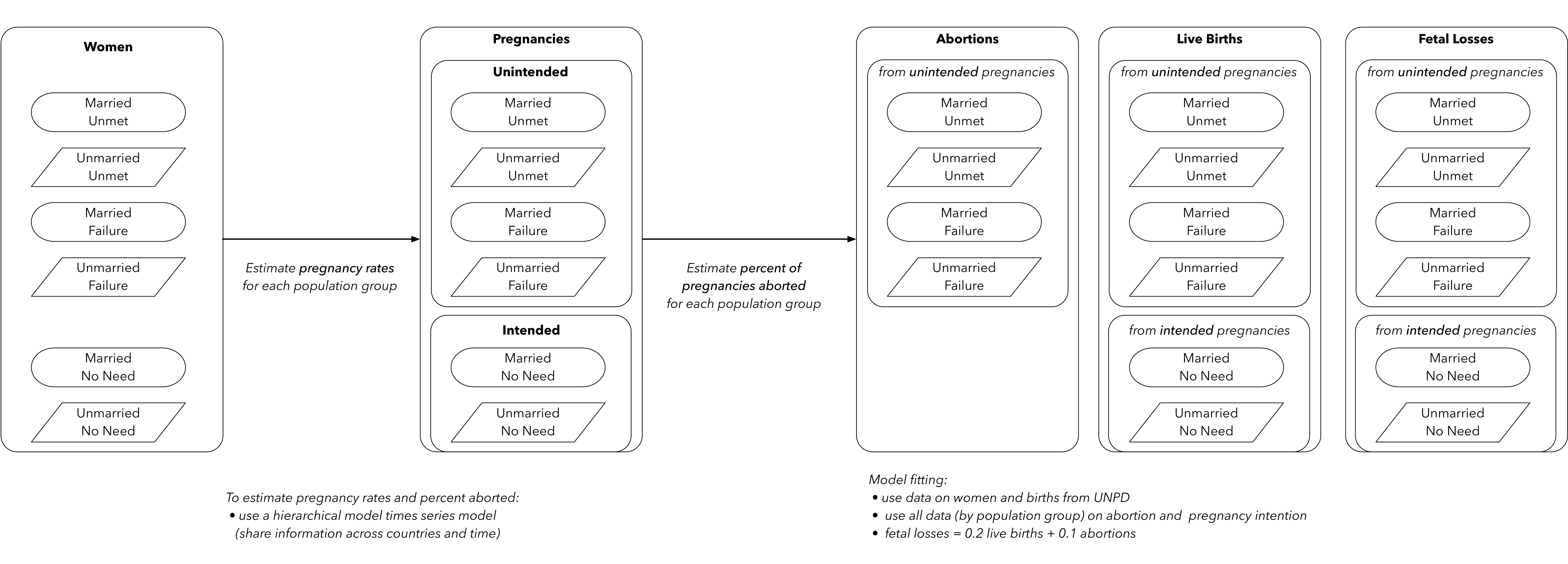}
    \caption{Model overview.}
    \label{fig-diagram}
\end{figure}

Our modeling strategy was informed by the proximate determinants of fertility. In brief, the incidence of unintended pregnancy is a function of the numbers of women with an unmet need for contraception and women using a contraceptive method who experience a method or user failure, separately by marital status, and the risk of pregnancy in each of these population groups (see Figure~\ref{fig-diagram}). Similarly, the incidence of intended pregnancy is a function of the number of women with no need for contraception, separately by marital status, and their risk of pregnancy.  

    As shown in Figure~\ref{fig-diagram}, intended pregnancies are attributed to married and unmarried women who want to have a(nother) child (``no need''), while unintended pregnancies are attributed to the other groups. These include those with an unmet need for contraception, as well as those with a met need for contraception, separate by marital status. In the diagram, rather than a ``met need'' group, we refer to women who experience a contraceptive ``failure'', as that is the subset of women who experience (unintended) pregnancy. 

	\label{unpd}Let $W_{c,t}$ denote the total number of women aged 15--49 for country $c$ and year $t$. We define population groups indexed by $f  = 1,\hdots, F$ where $F=7$, such that $W_{c,t} = \sum_{f=1}^F W_{f,c,t}$, where $W_{c,t,f}$ refers to the number of women in the subgroup indexed by $f$. Relevant groups, with non-zero pregnancy rates, are given by combining marital status with family planning information, as summarized in Table~\ref{tb:groups}. Group sizes $W_{c,t,f}$ are obtained from the  United  Nations Populations Division (UNPD),\footnote{United Nations, Department of Economic and Social Affairs, Population Division (2020). Estimates and Projections of Family Planning Indicators 2020.} combined with estimates of failure rates among modern users who are married.\footnote{Sedgh G, Bearak J, Singh S, Bankole A, Popinchalk A, Ganatra B, Rossier C, Gerdts C, Tun\c{c}alp \"{O}, Johnson BR, Johnston HB \& Alkema L. 2016. ``Induced abortion 1990 to 2014: Global, regional, and subregional trends.'' \textit{The Lancet.}} Note that, as per United Nations definitions and following the conventions within the literature, ``married'' includes (and ``unmarried'' excludes) all cohabiting women.

\newcommand{\unmarried}{\text{u}}
\newcommand{\married}{\text{m}}
\newcommand{\noneed}{\text{nn}}
\newcommand{\mnn}{\text{m.nn}} 
\newcommand{\munmet}{\text{m.unmet}} 
\newcommand{\mtrad}{\text{m.trad}}
\newcommand{\mfail}{\text{m.modfail}} 
\newcommand{\mmod}{\text{m.mod}}

\newcommand{\unn}{\text{u.nn}}
\newcommand{\uunmod}{\text{u.unmod}}
\newcommand{\umod}{\text{u.mod}}

\newcommand{\unintended}{\text{unintended}}
\newcommand{\intended}{\text{intended}}

\newcommand{\subsetofwomen}{\mathcal{F}}

\begin{table}[ht]
	\centering
	\begin{tabular}{|l|l|l|l|l|}
		\hline
	    marital status         & intention  & group name 1                  & group name 2        & abbreviation    \\ \hline 
		married (\married)     & intended   & no need for contraceptives    & no need             & $\mnn$          \\\cmidrule{2-5}
		                       & unintended & failures from modern methods  & mod-fail            & $\mfail$        \\
		                       &            & traditional methods           & trad                & $\mtrad$        \\
		                       &            & unmet need for any method     & unmet               & $\munmet$       \\ \hline 
		unmarried (\unmarried) & intended   & no need for contraceptives    & no need             & $\unn$          \\\cmidrule{2-5}
                               & unintended & modern users                  & modern              & $\umod$         \\
                               &            & unmet need for modern methods & unmod               & $\uunmod$       \\ \hline 
	\end{tabular}
	\caption{Grouping of women by marital and family planning status.}\label{tb:groups}
\end{table}

\subsection{Model summary}

The number of pregnancies $\npreg_{c,t}$ to occur in country $c$ during year $t$ is given by 
$$\npreg_{c,t} = \sum_{f=1}^F \npreg_{f,c,t},$$ 
where $\npreg_{f,c,t}$ refers to the number of pregnancies in population group $f$. 
The number of pregnancies to occur in a population group is in turn a function of the number of women in that group, $W_{f,c,t}$, and their risk of pregnancy, $\pregrate_{f,c,t}$:
$$\npreg_{f,c,t} = W_ {f,c,t} \cdot \pregrate_{f,c,t}.$$

The incidence of abortion within a population group, $\nabo_{f,c,t}$, is a function of the numbers of pregnancies in that group and the group-specific probability that a pregnancy will end in an abortion, the propensity to abort $\aboprop_{f,c,t}$:
$$\nabo_{f,ct} = \npreg_{f,c,t}\cdot \aboprop_{f,c,t}.$$
The incidence of abortion in a country-period is in turn the sum of the numbers of abortions across population groups, $\nabo_{c,t} = \sum_{f=1}^F \nabo_{f,c,t}$. 

Pregnancy outcomes are abortions, live births, and fetal losses (miscarriages and stillbirths). We estimated fetal losses using an approach derived from life tables of pregnancy loss by gestational age in which there is, on average, one fetal loss for every ten abortions, and one for every five live births. The resulting birth rate $\birthrate_{f,c,t}$ is given by 
$$\birthrate_{f,c,t} = \frac{\pregrate_{f,c,t}(1- 1.1 \aboprop_{f,c,t})}{1.2}.$$

We used hierarchical time series models to estimate latent pregnancy rates and propensities to abort by population subgroup. The next section provides a simplified illustration of the model structure and assumptions. The section subsequent to this then discusses how the subgroup parameters were modeled, followed by an explanation of the data and the data models, and the construction of aggregate outcomes. 

\subsection{Overview of model structure and assumptions}

\textit{Modeled parameters:} We modeled seven $\omega_{cpf}$, or pregnancy rate, parameters, for each country and five-year-period. However, we made simplifying assumptions to address identifiability challenges posed by data sparsity and measurement error.    For pregnancy parameters, we assumed for log-transformed rates:
\begin{enumerate}
    \item relative differences between countries within the same major cluster were modeled for all married contraceptive users collectively (such that differences from the regional means were equal  for log-transformed $\omega_{\mfail}$ and $\omega_{\mtrad}$), 
    \item country trends for log-transformed $\omega_{\uunmod}$ followed those for log-transformed $\omega_{\umod}$. 
    \item there is no time trend for $\omega_\unn$, 
    \item  pregnancy rates were lowest for $\omega_\unn$, which includes sexually inactive women, and highest for $\omega_\mfail$.
\end{enumerate}

  Abortion rates $\alpha_{cpf}$ varied by marital status but we assumed that changes between any two adjacent five-year periods were equal for approximately logit-transformed (see below) $\alpha_\married$ and $\alpha_\unmarried$,

\textit{Groupings of countries:} To exchange information on the group-specific pregnancy rates and propensities to abort within the model, we developed a four-level hierarchy in which countries were grouped according to expected relationships between the predictors and the outcomes. Countries were nested within regions and super-regions, and the parameters for the super-regions were modeled around global means.

To develop the 4-level hierarchy, we compared the 3-level groupings of countries used in previous abortion and unintended pregnancy studies to the groupings used in the Global Burden of Disease (GBD) Study. We largely adopted the GBD's ``super regions'' as our major clusters, with the exception that we created a ``South Asia, Southeast Asia and Oceania'' major cluster, retaining the ``Viet-China'' region used in previous abortion studies, whereas GBD has ``South Asia'' and ``Southeast Asia, East Asia, and Oceania.'' We also found that the ``Revised Eastern Europe'' region used in previous studies corresponded to the ``Central Europe, Eastern Europe and Central Asia'' super-region. However, whereas in the GBD this super-region contains three regions, ``Central Europe'', ``Central Asia,'' and ``Eastern Europe'', due to extreme heterogeneity, we made seven sub-clusters within this cluster: ``Central Europe'', ``Baltics'', ``Balkans'', ``Central Asia'', ``Eastern Asia'' (Mongolia), ``Caucasus'', and ``Eastern Europe''. Other major clusters correspond those used in the GBD -- ``High-income'', ``Latin America and the Caribbean'', ``Sub-Saharan Africa'', and ``Middle East and North Africa''. A complete list of countries is in Bearak et al. (2020). 

\subsection{Modeling subgroup-specific pregnancy rates}
Given data sparsity, subgroup-specific pregnancy rates were estimated (i) by 5-year period, (ii) using hierarchical models to exchange information across countries with regions, (iii) using time series processes to exchange information across periods and (iv) using substantive information on subgroup characteristics to incorporate relative constrains across groups. With slight misuse of notation for simplification, in the remaining text, the unknown parameters  $\pregrate_{f,c,p}$ refer to 5-year period $p$ for subgroup $f$ (see Table~\ref{tb:groups} for information on groupings). 

We first explain the specification of the parameters for reference period $p_0$, here 2000-2004, followed by the specification for other periods, and finally, the set-up for non-country-specific parameters.

\subsubsection{Reference period}
In reference period $p=p_0$, hierarchical distributions are used for log-transformed pregnancy rates for groups $f \in \{f_{\mnn}, f_{\unn}, f_{\munmet}, f_{\mfail}, f_{\uunmod}\}$:
$$\log\pregrate_{c,p,f} \sim t_3\left(
\log\pregrate\region_{r[c],p,f},
\sigma_{\pregrate, f}^2
\right) \:
T\left(
\text{lower bound}, 
\text{upper bound}
\right),$$
where $\pregrate\region_{r[c],p,f}$ refers to the mean pregnancy rate in the region $r[c]$ that country $c$ belongs to,
$\sigma_{\pregrate,f}^2$ the across-country variance in pregnancy rates, and lower and upper bounds defined in Table~\ref{tab-boundspreg}. The bounds are used to (i) impose minima and maxima rates for the married-no-need group, (ii) to impose relative constraints between the marital groups based on characteristics of the unmarried group (unmarried women includes sexually inactive women), (iii) to constrain the pregnancy rate among failures to be higher than the other rates. 

\begin{table}[ht]
	\centering
	\begin{tabular}{|l|l|l|}
		\hline
	subgroup & lower bound & upper bound \\ \hline
		married, intended & $\log\pregrate_{\text{unmarried.no.need}}$ & $\log(1.1)$ \\
		unmarried, intended & $\log(0.001)$ & $\log\pregrate_{\text{married.no.need}}$ \\
		married, unmet for modern & $\log\pregrate_{\text{unmarried.no.need}}$ & $\log(1.1)$ \\
		unmarried, unmet for modern & $\log\pregrate_{\text{unmarried.no.need}}$ & $\log(1.1)$ \\
		married, failures from modern use & max among all groups & $\log(5)$ \\ \hline
	\end{tabular}
	\caption{Bounds used for pregnancy rates in reference period $p_0$.}\label{tab-boundspreg}
\end{table}

The rate for unmarried-modern, \umod, is defined relative to the rate for unmarried women with an unmet need for modern methods (\uunmod):
$$\log\pregrate_{c, p_0, \umod}=
\log\pregrate_{c,p_0, \uunmod} +
\gamma_{c,\umod},$$
where 
$$\gamma_{c,\umod}  \sim 
t_{3}\left(\gamma_{r[c],\umod}, \sigma^2_{\pregrate, \umod}\right) \:
T\left(\log\omega_{c,p,\unn} - \log\omega_{c,p,\uunmod}, 0 \right)$$
such that this rate is greater than the unmarried-no-need $\unn$ rate and does not exceed the unmarried-unmet-need-for-modern $\uunmod$ rate. 

The rate for married women using traditional methods is defined by a regional rate, which is multiplied by the country-specific ratio of the married-mod-fail to super-regional married-mod-fail pregnancy rates:	
$$\pregrate_{\mtrad, c, p_0} = \pregrate_{\mtrad,q[c],p_0}\cdot\frac{\pregrate_{\mfail,c,p_0}}{\pregrate_{\mfail,q[c],p_0}},$$
where $q[c]$ refers to the super-region of country $c$. 

\subsubsection{Time trends}
For period $p$ before or after $p_0$, with neighboring period $p^*[p] = p-1$ for $p>p_0$ and $p^*[p] = p+1$ for $p<p_0$, we capture the relative change between $p^*[p]$ and $p$ on the country level through additions to log-transformed pregnancy rates. The implementation follows that of random walks, with regional drift terms.
In the parameterization that follows, $\Delta_{c,p,f} = \pregrate_{c,p,f} - \pregrate_{c,p^*[p],f}$ and $\Delta\region_{c,p,f} = \pregrate\region_{c,p,f} - \pregrate\region_{c,p^*[p],f}$. 

	For intended pregnancies among married women, country-specific random walks are defined as follows: 
	
$$\log\pregrate_{c,p,\mnn} \sim t_{3}(\log\omega_{c,p^*[p],\mnn} + \Delta\region_{r[c],p,\mnn}, \sigma^2_{\delta,\mnn})T(\log \pregrate_{c,p, \unn}, \log(1.1)),$$

where $\Delta\region_{r[c],p,\mnn}$ is the regional estimate for the change from period $p^*[p]$ to $p$ and  $\sigma^2_{\delta,\mnn}$ the variance associated with the change. Bounds are used to impose the same constraint as in the reference period such that the pregnancy rate is lowest among unmarried-noneed women.

For intended pregnancies among unmarried women, we do not estimate a trend:

$$\log\pregrate_{c,p,\unn} = \log\pregrate_{c,p^*[p],\unn}.$$

Trends in unintended pregnancies to married women with unmet need for any method are modeled relative to trends in intended pregnancies among married women. 
\begin{eqnarray*}
	\log\pregrate_{c,p,\munmet} &\sim& N(\log\pregrate_{c,p^*[p],\munmet} + (\Delta_{c,p,\mnn} - \Delta\region_{r[c],p,\mnn}) + \Delta\region_{r[c],p,\munmet},\\
	&& \sigma^2_{\delta,\munmet})T(\log\pregrate_{c,p,\unn} , \log(1.1)).
\end{eqnarray*}

Trends in failures from modern methods among married women are modeled relative to trends in unintended pregnancies to married women with unmet need for any method:
\begin{eqnarray*}
	\log\pregrate_{c,p,\mfail} &\sim& t_{3}(\log\pregrate_{c,p^*[p],\mfail} + \Delta_{c,p,\munmet}, \\
	&&\sigma^2_{\delta,\mfail})T(m_{c,p,\mfail}, \log(5)),
\end{eqnarray*}
where $ m_{c,p,\mfail} = max( \log \pregrate_{c,p, \munmet}, \log \pregrate_{c,p, \mnn}, \log \pregrate_{c,p, \uunmod})$ to constrain failures to have the highest rates in all periods.

For  traditional method users, we applied the same trends as for failures, i.e., 
\begin{eqnarray*}
	\log\pregrate_{c,p,trad} &=& \log\pregrate_{c,p^*[p],trad} + \Delta_{c,p,\mfail}.
\end{eqnarray*}

For unmarried women, unintended pregnancies among women with an unmet need for modern methods, we set
\begin{eqnarray*}
	\log\pregrate_{c,p,\uunmod} &\sim& t_{3}(
	\log\pregrate_{c,p^*[p],\uunmod} + \Delta\region_{r[c],p^*[p],\uunmod}, \sigma^2_{\delta,\uunmod})T(\\
	&&\log\pregrate_{c,p,\unn},
	\log(1.1)).
\end{eqnarray*}

For unintended pregnancies among unmarried modern users, relative changes are assumed to be the same as those of the unmarried-unmet-need group:
\begin{eqnarray*}
	\log\pregrate_{c,p,\umod} &=& 
	\log\pregrate_{c,p^*[p],\umod} + \Delta_{c,p,\uunmod}.
\end{eqnarray*}

	\subsection{Modeling subgroup-specific propensities to abort}
\newcommand{\laboprop}{\lambda} 

Propensity to abort was estimated by 5-year period, separately for married women with unintended pregnancies and unmarried women with unintended pregnancies. Again, with slight misuse of notation for simplification, in the remaining text, the unknown parameters $\aboprop_{c,p,f}$ refer to 5-year period $p$ for the two subgroups married-unintended and unmarried-unintended. 

The abortion parameters are modeled on a transformed scale to constrain $0 < \alpha < 1/1.1$, where the upper bound is introduced to guarantee positive pregnancy rates after accounting for 0.1 stillbirth being associated with every abortion. Let $\laboprop = \frac{1/1.1}{1+\exp(-\alpha)}$, the transformed probability to abort and $\Lambda$ the period-change. For abortions among unintended pregnancies to married women:	
\begin{eqnarray*}
	\laboprop_{c,p_0, \married} &\sim& N(\laboprop^{reg}_{r[c],p_0,\married}, \sigma^2_{\lambda,\married})T(10^{-6}, 1-10^{-6}),\\
	\laboprop_{c,p, \married} &\sim& N(\laboprop_{c,p^*[p],\married} + \Lambda\region_{r[c],p,\married}, \sigma^2_{\Lambda,\married})T(10^{-6}, 1- 10^{-6}), \text{ for }p \neq p_0.
\end{eqnarray*}

Abortions among unintended pregnancies among unmarried women are modeled relative to those among married women:
	\begin{eqnarray*}
	\laboprop_{c,p_0, \unmarried} &\sim& N(\laboprop_{c,p_0, \married} + \laboprop\region_{r[c],p_0,\unmarried} - \laboprop\region_{r[c],p_0,\married}
	, \sigma^2_{\laboprop,\unmarried})T(10^{-6}, 1-10^{-6}),\\
	\laboprop_{c,p, \unmarried} &=& \laboprop_{c,p^*[p],\unmarried} + 
	\Lambda_{c,p, m}, \text{ for }p \neq p_0.
\end{eqnarray*}

\subsection{Non-country specific parameters}
\subsubsection{Regional and global mean parameters}
At the regional level, referring to $r[c]$, the same process as described above at the country-level, with hierarchical means replaced by super-region means and variances by regional variance parameters. For example, for married-intended, the random walk is given by
$$\log\pregrate\region_{r,p,\mnn} \sim N(\log\omega_{r,p^*[p],\mnn} + \Delta\superregion_{q[r],p,\mnn}, \sigma^{(region)2}_{\delta,\mnn})T(\log \pregrate_{r,p, \unn}, \log(1.1)),$$
where $\Delta\superregion_{q[r],p,\mnn}$ is the regional estimate for the change from period $p^*[p]$ to $p$ and  $\sigma^{(region)2}_{\delta,\mnn}$ the variance associated with the change. At the super-region-level, again the same process is used with variance parameters equal to those at the regional level, and world-level mean drifts. For traditional use at the super-regional level,
$$\log \omega_{q,p_0, \mtrad} \sim N(\log \omega\world_{\mtrad}, \sigma^{(region)2}_{\omega, \mtrad})T(log(.061),),$$
where the lower bound is set based on a published analysis of DHS data;\footnote{Polis CB et al., \textit{Contraceptive Failure Rates in the Developing World: An Analysis of Demo- graphic and Health Survey Data in 43 Countries}, New York: Guttmacher Institute, 2016, http://www.guttmacher.org/ report/contraceptive-failure-rates-in-developing-world} as abortion is known to be under-reported in those data, the true rate is expected to be no less than this value. The corresponding global mean, $\log\omega\world _{p_0, \mtrad}$, defined below, is informed by additional research on the average failure rate for traditional method use.\footnote{Hatcher RA, et al., Contraceptive Technology. 20 ed. 2011}

	At the global level, pregnancy rates in $p_0$ are defined as follows:
\begin{eqnarray*}	
\omega\world_{p_0, \mnn} &\sim& U(\log(.001), \log(1.1)),\\
\omega\world_{p_0, \unn} &\sim& U(\log(.001), 	\omega\world_{p_0, \mnn} ),\\
\omega\world_{p_0, \mnn} &\sim& U(\omega\world_{p_0, \unn}, \log(1.1)),\\
\omega\world_{p_0, \umod} &\sim& U(\omega\world_{p_0, \unn}, \omega\world_{p_0, \uunmod}),\\
	\omega\world_{p_0, \mtrad} &\sim& U(\omega\world_{p_0, \unn}, \log(1.1)),\\
\log\omega\world _{p_0, \mtrad} &\sim& N(\log(.23), 0.5^2)T(log(.061), \log \omega_{p_0, \munmet}).
\end{eqnarray*}	
For other periods and $f \neq f_{mnn}$, all rates are kept constant: $\omega\world_{p, f} = \omega\world_{p_0, f}$. For the married-noneed group for $p \neq p_0$, a random walk is used: 
\begin{eqnarray*}	
\log \omega\world_{p, \mnn} &\sim& N(\log \omega\world_{p^*[p], \mnn}, \sigma^{(region)2}_{\delta, \mnn})T(\log \omega\world_{p, \unn}, \log(1.1)).
\end{eqnarray*}

For regional abortion parameters $\laboprop\region_{r[c],p,\married}$ and $\laboprop\region_{r[c],p,\unmarried}$, 

\begin{eqnarray*}
	\laboprop\region_{r,p_0,  \married} &\sim& N(\laboprop\superregion_{q[r],p_0,\married}, \sigma^{(region)2}_{\lambda,\married})T(10^{-6}, 1-10^{-6}),\\
	\laboprop\region_{r,p,    \married} &\sim& N(\laboprop\superregion_{r,p^*[p],\married} + \Lambda\superregion_{q[r],p,\married}, \sigma^{(region)2}_{\lambda,\married})T(10^{-6}, 1- 10^{-6}), \text{ for } p \neq p_0,\\
	\laboprop\region_{r,p_0,\unmarried} &\sim& N(\laboprop\region_{r,p_0,\married} + \laboprop\superregion_{q[r],p_0,\unmarried} - \laboprop\superregion_{q[r],p_0,\married}, \sigma^{(region)2}_{\lambda,\unmarried})T(10^{-6}, 1-10^{-6}).
\end{eqnarray*}

At the global level, propensities to abort are defined as follows:

\begin{eqnarray*}	
\alpha\world_{m}         &\sim& U(10^{-6}, 1- 10^{-6}),\\
\laboprop\world_{p_0, m} &=&    \frac{1}{1+\exp(-\alpha\world_{m})},\\
\laboprop\world_{p_0, u} &\sim& N(\laboprop\world_{p_0, m}, 0.5^2)T(10^{-6}, (1 - 10^{-6}),\\
\laboprop\world_{p,m} &=& \laboprop\world_{p_0, m} \text { for } p \neq p_0,\\
\laboprop\world_{p,u} &=& \laboprop\world_{p_0, u} \text { for } p \neq p_0.
\end{eqnarray*}

	\subsubsection{Variance parameters}
For standard deviations $\sigma_{\omega, f}$, the following default priors are used: 
\begin{eqnarray*}
	\sigma_{\omega, f} &\sim& N(.01, 0.5^2)T(.01, \sigma\region_{\omega, f}),
\end{eqnarray*}
where $\sigma\region_{\omega, f}$ refers to the regional standard deviation for the respective parameter. Standard deviations were set equal for pregnancy rate parameters for subgroups $f_{\mfail} = f_{\munmet}$ and $f_{\uunmod} = f_{\unn}$.

For standard deviations for time trends, the country standard deviations are equal to the regional standard deviations:
\begin{eqnarray*}
\sigma_{\delta, f} &=& \sigma\region_{\delta, f}.
\end{eqnarray*}

For abortions, the same default priors are used with context specific upper bounds:
\begin{eqnarray*}
\sigma_{\lambda, m} &\sim& N(.001, 0.5^2)T(.001, \sigma\region_{\lambda, m}),\\
\sigma_{\lambda, u} &\sim& N(.001, 0.5^2)T(.001, \max(\sigma\region_{\lambda, u}, \sigma_{\lambda, m})),\\
	\sigma_{\Lambda, m} &\sim& N(.001, 0.5^2)T(.001, \sigma\region_{\Lambda, m}).
\end{eqnarray*}

At the regional and super-regional level, variance parameters are defined in a similar manner. 	For standard deviations $\sigma\region_{\omega, f}$, the following default priors are used: 
\begin{eqnarray*}
\sigma\region_{\omega, f} &\sim& N(.001, 0.5^2)T(.001, 3),
\end{eqnarray*}
where $\sigma\region_{\omega, f}$ refers to the regional standard deviation for the respective parameter. Standard deviations were set equal for pregnancy rate parameters for subgroups $f_{\mfail} = f_{\munmet}$, $f_{uunmod} = f_{umod} = f_{mtrad}$.

For standard deviations for time trends $\sigma_{\delta, f}$, we set
\begin{eqnarray*}
\sigma\region_{\delta, f} &\sim& N(.01, 0.5^2)T(.01, 1),
\end{eqnarray*}
with standard deviations set equal for $f_{\mnn} = f_{\unn} = f_{\munmet}$, and constrained so that $f_{\mfail} \leq \min(f_\mnn, f_\munmet)$.

Finally, for abortions,
\begin{eqnarray*}
\sigma\region_{\lambda, m} &\sim& N(.001, 0.5^2)T(.001, 3),\\
\sigma\region_{\lambda, u} &\sim& N(.001, 0.5^2)T(.001, \sigma\region_{\lambda, m}),\\
\sigma\region_{\Lambda, m} &\sim& N(.001, 0.5^2)T(.001, \sigma\region_{\lambda, m}).
\end{eqnarray*}

\subsection{Data models}	

\newcommand{\datavar}{\zeta}
\newcommand{\obsprop}{y}
\newcommand{\estprop}{\Xi} 
\newcommand{\sdy}{s}
\subsubsection{Birth rates}

Data on the numbers of births in each country-period, $B_{cp}$, were provided by the United Nations Population Division (UNPD), with
$$\textit{log }B_{cp} \sim N(\textit{log }\nbirths_{cp}, .025^2)$$
for countries which, according to the UNPD's database, had high-quality vital registration systems, and otherwise $\textit{log }B_{cp} \sim N(\textit{log }\nbirths_{cp}, .05^2)$.

\subsubsection{Data on the proportions of pregnancies and births unintended}

All data on observed proportions were combined across types (i.e. unintended pregnancies or births), across countries and indexed by $i = 1, 2, \hdots, I$; for proportion $\obsprop_{i}$, $c[i]$ refers to the country of the observation, $t_1[i], \hdots , t_{n[i]}$
refer to the calendar years of the observation period and $\subsetofwomen[i]$ denotes the subgroup that the observation refers to (which may be all women $\subsetofwomen[i] = \subsetofwomen^{all}$, married women only $\subsetofwomen[i] = \{\munmet, \mtrad, \mfail\}$, or unmarried women only $\subsetofwomen[i] = \{\uunmod, \umod\}$).

For data on the proportion of pregnancies or births unintended, we assume that the mode of the sampling model for the data is given by the true (modeled) proportion as follows:

\begin{eqnarray}
\obsprop_i &\sim& N\left(\frac{\estprop_i}{\mu_{d,q[i]}} , \datavar_i\right)T(0,1),\label{eq-ub}
\end{eqnarray}
where mean $\estprop_i$ is the true (modeled) proportion for the corresponding type-country-year-group combination, and $\mu_{qd}$ is a super-region-specific bias term for data on pregnancies or births that occurred up to $d = 3$ or $d = 5$ years before the date a respondent was interviewed in super region $q$. 

The modeled proportion is given by
\begin{eqnarray*}
\estprop_i &=& \frac{\sum_{t=t_1[i]}^{t_{n[i]}}\sum_{f in \subsetofwomen[i]^{\text(unintended)}} \nbirths_{c,t,f}}{\sum_{t=t_1[i]}^{t_I[i]}\sum_{f in \subsetofwomen[i]}\nbirths_{c,t,f}}, \text{ for data on proportions of births unintended, and } \\
\estprop_i &=& \frac{\sum_{t=t_1[i]}^{t_{n[i]}}\sum_{f in \subsetofwomen[i]^{\text(unintended)}} \npreg_{c,t,f}}{\sum_{t=t_1[i]}^{t_I[i]}\sum_{f in \subsetofwomen[i]}\npreg_{c,t,f}}, \text{ for data on proportions of pregnancies unintended.}
\end{eqnarray*}

The bias term is modeled,
\begin{eqnarray*}
\mu_{q,1} &=& 1,\\
\log(\mu_{q,3}) &\sim&  N(u_{3}, \sdy^2_{3})T(0, \mu_{q,5}),\\
\log(\mu_{q,5}) &\sim& N(u_{5}, \sdy^2_{5})T(0, ),
\end{eqnarray*}
with mean $u_d$ and standard error $\sdy_d$ taken from a multi-level model.\footnote{This is estimated with a multi-level model using micro-data from the Demographic and Health Surveys and the Multiple Indicator Cluster surveys, comparing responses about births in the year prior to interview to responses 3 or 5 years prior to interview.}

Variance $\datavar_i$ is the sum of fixed and unknown variances which are defined as follows:
\begin{equation}\label{eq-totalvar-ub}
\datavar_i = \sdy_i^2 + (\errormult^{(non-sampling)}_{\subsetofwomen[i]}+r[i]\cdot \errormult^{(non-representative)})y_i(1 - y_i),
\end{equation}
where $\sdy_i^2$ is the sampling error for the observed proportion, $\errormult^{(non-sampling)}$ is an unknown multiplied by the Bernoulli variance, $y_i(1 - y_i)$, to account for non-sampling error; and $r[i]=1$ if the proportion was computed from a non-representative survey and 0 if the survey was representative, such that $\errormult^{(non-representative)}$ allows for additional uncertainty when incorporating proportions computed from non-representative surveys. 
Sampling error $\sdy_i^2$ is calculated to take into account sample size and survey design
for those surveys where microdata were available. For other surveys, it is given by the product of the binomial variance and an inflation factor, computed as the ratio of the design-adjusted variance and the binomial variance from surveys where microdata were available

For observed proportions that are to be treated as ranges, indexed by $i=I+1, \hdots , I+N$, notation is the same as described above and subscripts (min) and (max) denote the lower and upper bounds. These data were used to constrain the true (modeled) proportions that they refer to as follows:  
\begin{eqnarray}
\obsprop^{*(min)}_i \leq & \estprop_i & \leq \obsprop^{*(max)}_i, 
\end{eqnarray}
where $\obsprop^{*(min)}_i$ and $\obsprop^{*(max)}_i$ account for uncertainty in the ranges:
\begin{eqnarray*}
	\obsprop^{*(min)}_i & \sim & N\left(\obsprop^{(min)}_i,\datavar_i^{(min)}\right)T(0, 1), \\
	\obsprop^{*(max)}_i & \sim & N\left(\obsprop^{(max)}_i,\datavar_i^{(max)}\right)T(\obsprop^{*(min)}_i, 1),
\end{eqnarray*}
where $\datavar_i^{(min)}$ and $\datavar_i^{(max)}$ are given by Eq.~\ref{eq-totalvar-ub}, replacing $\obsprop_i$ by $\obsprop^{(min)}_i$ and $\obsprop^{(max)}_i$. 

Finally, the variance multipliers were given diffuse priors, with the non-sampling variance multipliers constrained such that there is one multiplier for data on the proportion of births or pregnancies unintended (i.e., where $f = \subsetofwomen^{all}$), and another for data on the percent distribution by subgroup.
\begin{eqnarray*}
	\sqrt{\errormult^{(non-sampling)}_{\subsetofwomen^{all}}} & \sim & N(.01, .5^2)T(.01, 4), \\
	\sqrt{\errormult^{(non-sampling)}_{f}} & \sim & N(.01, .5^2)T(.01, 4), \text{ for } f = f_{married}, f_{unmarried}, \\
	\sqrt{\errormult^{(non-representative)}} & \sim & N(.01, .5^2)T(.01, 4).
\end{eqnarray*}

\subsubsection{Data on abortion incidence}

Observations of abortion counts were combined across countries and indexed by $j = 1, 2, \hdots, J$; for count $a_{j}$, $c[j]$ refers to the country of the observation, $p[j]$ refers to the observation period and $\subsetofwomen[j]$ denotes the subgroup that the observation refers to (which may be all women $\subsetofwomen[j] = \subsetofwomen^{all}$, or married women only $\subsetofwomen[j] = \{\munmet, \mtrad, \mfail\}$).

For abortion incidence count data, we assume:
\begin{eqnarray}
\log(a_j) &\sim& N(\log(\nabo_{c[j],p[j],\subsetofwomen[j]}\cdot \beta_j),\iota_j^2 + \iota_j^2  e_j  \pi ),\label{eq:dm_abo}
\end{eqnarray}
where $a_{j}$ is the $j$-th observed abortion count for country $c[j]$ in period $p[j]$ among women in subgroup(s) $f \in \subsetofwomen[j]$, $\nabo_{c[j],p[j],\subsetofwomen[j]}$ is the true (modeled) abortion count for the corresponding country-period-group combination, $\beta_j$ is a bias term, the standard error $\iota_j$ depends on the data source. Finally, $e_j = 1$ if the datum comes from a study which employed a non-representative study with $\pi \sim N(.01, 4)T(.01, 4)$. $e_j=0$ for all other studies, including official statisics.

\paragraph{Official statistics} For official statistics including minima, $\beta_j = 1$ and $\iota_j = \iota^{(region)}_{r[j]}$, with
\begin{eqnarray*}
\iota_r^{(region)} \sim & N(.025, 4)T(.025, 1), & \text{ for } q[r] = q^{\text{high income cluster}}, \\
\iota_r^{(region)} = & \iota^{\text{other clusters}}, & \text{ for } q[r] \ne q^{\text{high income cluster}}, \\
\iota^{\text{other clusters}} \sim & N(.025, 4)T(.025, 1).
\end{eqnarray*}

For observed counts that are to be treated as a minimum or maximum, indexed by $j = J+1, \hdots, J+K$, these data were used to constrain the true (modeled) counts that they refer to as follows:
\begin{eqnarray*}
\log(a^*_j) \leq & \log(\nabo_{c[j],p[j],\subsetofwomen[j]})_j, \text{ for data treated as min,}\\
\log(a^*_j) \geq & \log(\nabo_{c[j],p[j],\subsetofwomen[j]})_j, \text{ for data treated as max,}
\end{eqnarray*}
where $a^*_i$ accounts for uncertainty in the observation:
\begin{eqnarray*}
	\log(a^*_j) & \sim & N(\log(a_j)),\iota_j^2).
\end{eqnarray*}

For observed counts that are to be treated as ranges, indexed by $j = J + K + 1, \hdots, J+K+L$:

\begin{eqnarray*}
	\log(a^{*(max)}_j) & \sim & N(a^{(max)}_j,\iota_j^2),\\
	\log(a^{*(min)}_j) & \sim & N(a^{(max)}_j,\iota_j^2)T(, \log(a^{*(max)}_j)).\\
\end{eqnarray*}

\paragraph{Surveys} For surveys of women from Central Europe, Eastern Europe, Central Asia and the Caucasus, which were not treated as minima, data are included as per Eq.\ref{eq:dm_abo} with sampling error $\iota_j = \iota^{(survey)} \sim N(.05, 4)T(.05, 1)$ and bias 
$$\beta_j \sim N(.65, \iota^{(survey bias)2}T(.0475, 1),$$ 
with $\iota^{survey bias} ~ \sim N(.125, 4)(.125, 1)$, reflecting that, on average, women in those regions report about two-thirds of abortions. The same average adjustment was used in previous studies of abortion incidence,\footnote{Sedgh et al 2016; Sedgh et al 2012; Sedgh et al 2007} except that this was computed from a multi-level model, from which the standard error was taken.

\paragraph{Data from indirect methods} For abortion counts from published studies using an indirect method,  $a_j$ does not correspond to the study's estimate of the number of abortions, but rather, the number of abortions estimated from the direct component of the study. We assume

\begin{eqnarray*}
\log(\beta_j) \sim N(\log(b_j), \eta^{2(indirect)}),
\end{eqnarray*}

where $b_j$ is taken from the study and  $\eta^{(indirect)} \sim U(.1, .2)$ to allow for additional uncertainty attributable to the indirect component of the study and $\iota_j = \eta^{(direct)}$, with:
\begin{eqnarray*}
\eta^{(direct)} \sim N(.05, \eta^{(indirect)}).
\end{eqnarray*}

\paragraph{Abortions among married women}
Observations of the proportion of abortions that occurred to married women were combined across countries and indexed by $h=1,2,\hdots , H$, where  $c[h]$ and $t[h]$ refer to the country and calendar year of the observation and observation $y_h$ refers to the proportion of abortions to occur to married women:

\begin{eqnarray*}
	m_{h} \sim N\left(y^*_h,\iota^2_{(prop)}\right),
\end{eqnarray*}

where $m_{h}$ is the $h$-th observed proportion of abortions among married women for country $c[h]$ in year $t[h]$, $m^*_h$ is the modeled proportion, $m^*_h = \frac{\sum_{f \in married}\nabo_{c[h],t[h],f}}{\nabo_{c[h],t[h]}}$ and standard error $\iota_{(prop)} \sim N(.01, .5^2)T(.01, 4)$.

These data were generally treated as minima or maxima, in which case they were modeled,

\begin{eqnarray*}
	m_{h}^{*} &\sim& N(m_{h}, {\iota}^2_{(prop)})T(0,1) \text{, with} \\
    m_h &\geq& m_h^*, \text{ if observation } h \text{ refers to a minimum , and }\\
	m_h &\leq& m_h^*, \text{ if observation } h \text{ refers to a maximum}.
\end{eqnarray*}

\subsection{Reported estimates}

The unintended pregnancy and abortion model is a Bayesian model whereby interest lies in the posterior distributions of the outcomes of interest, which summarizes all available information about the outcome of interest. We used a Markov Chain Monte Carlo (MCMC) algorithm to generate samples of the posterior distributions of all model parameters. The MCMC sampling algorithm was implemented using JAGS 4.3.0 open source software \footnote{Plummer, Martyn. JAGS: A Program for Analysis of Bayesian Graphical Models Using Gibbs Sampling (version 4.3.0), 2017. https://sourceforge.net/projects/mcmc-jags.
} and the analysis was carried out in R 3.6.2 \footnote{R Core Team. R: A Language and Environment for Statistical Computing. https://www.R-project.org/}. The sampling algorithm produced a set of trajectories of unintended pregnancy, unintended birth, and proportion of unintended pregnancies that end in abortion for each country, from which estimates for groups of countries were derived.

We computed point estimates using the posterior medians of the estimates for each country and for each grouping of countries, separately for each five-year period.  We computed 80\% uncertainty intervals (UIs) using the 10th and 90th percentiles of the posterior distributions. The interpretation of such intervals is that there is a 10\% chance that the true outcome is below the interval, and there is a 10\% chance that the true outcome is above the interval.

\subsection{Summary of major differences compared to past approaches}
\label{sec:diffs}

\subsubsection{Process model}
Our statistical model closely corresponds to our theoretical framework, in contrast to earlier studies. Family planning indicators predict unintended pregnancy, but Sedgh and colleagues used these to estimate abortion. Bearak and colleagues previously produced unintended pregnancy estimates informed by this theoretical framework, but, in their study, the unknowns to be modeled were the percentages of births unintended; the abortion estimates from Sedgh et al were treated as fixed. This means that in prior studies, pregnancy intention data did not inform the previously published abortion estimates, and uncertainty in the abortion estimates was ignored when modeling unintended births. In contrast, our model uses a joint-estimation approach, simultaneously producing estimates for all outcomes that are for the first time informed by all available data.

\subsubsection{Family planning indicators}
\begin{figure}[htb!]
    \centering
    \includegraphics[scale=0.27]{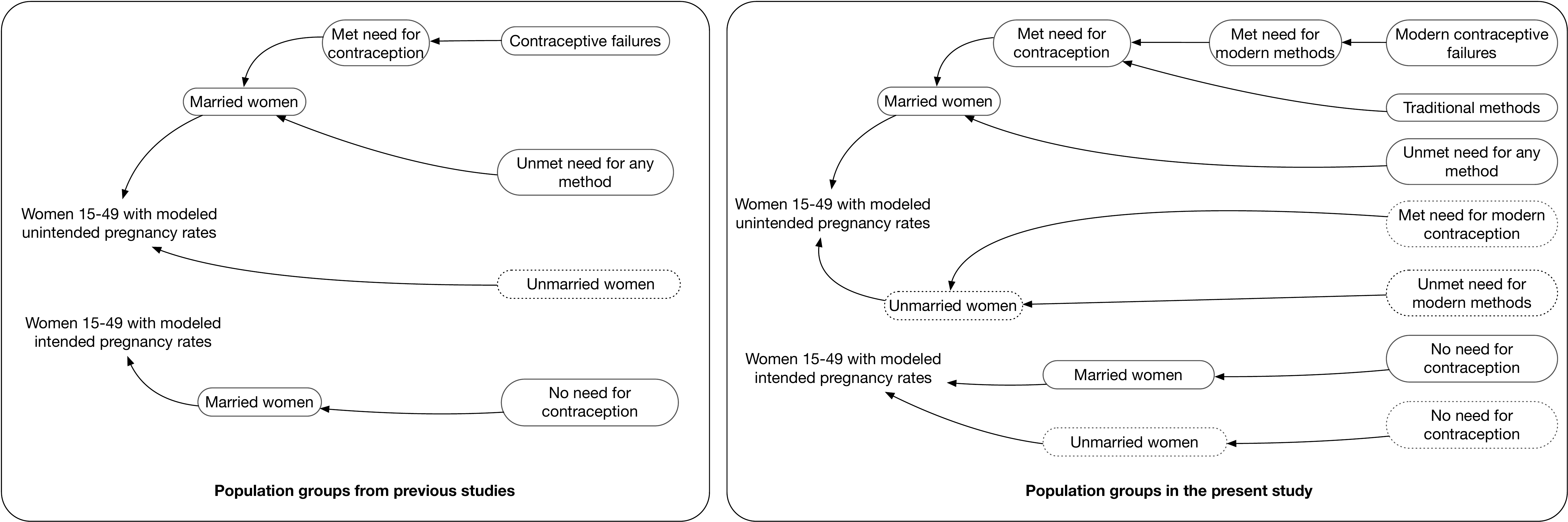}
    \caption{Population groups used in prior studies (left) and current study (right).}
    \label{fig-groups}
\end{figure}

Our statistical model uses information on sexual activity, contraceptive needs and use among unmarried women, whereas earlier studies did not, see Figure~\ref{fig-groups}. This was made possible by the UNPD's recently published estimates of contraceptive need and use among unmarried women. As a result of this difference, our model estimates unintended pregnancies in proportion to sexually active unmarried women, rather than all unmarried women.

We also distinguish between modern and traditional method use among married women differently as compared to previous studies. We follow the procedure first outlined by Sedgh and colleagues to estimate failure rates among married modern users. However, previous studies computed a ``contraceptive failures'' group by summing $W_\mfail + W_\mtrad\cdot \omega_\mtrad$ with $\omega_\mtrad = .23$, whereas we treated $\omega_\mtrad$ as an unknown which varied across super-region.

\subsubsection{Groupings of countries}
To exchange information across countries on the group-specific pregnancy rates and propensities to abort within the model, we developed a four-level hierarchy. This contrasts with the approach taken by earlier studies, in which three-level hierarchies were used, and subregions were collapsed based on data availability. As a result of that approach, the hierarchy used to model abortions differed from the hierarchy used to estimate unintended pregnancies (with Sub-Saharan Africa being one region in Sedgh et al, and two regions in Bearak et al; and with Western Europe being three regions in Sedgh et al, and one region in Bearak et al). These differences were necessary for model identifiability in those studies. In contrast, with our four-level hierarchy, we were able to use the same groupings for all outcomes and did not have to collapse sub-regions together.

\subsubsection{Data collection}
We built a database that is substantially larger compared to that used by previous studies. Sedgh et al previously obtained abortion data for 92 countries, whereas we obtained abortion data for 104 countries. Bearak et al previously obtained intention data for 105 countries, whereas we obtained intention data from 139 countries. In total, our database includes abortion and intention data from 166 countries in total.

Our data search processes were similar. However, we processed micro-data from the Multiple Indicator Cluster Surveys, whereas previous studies did not. We also conducted an official World Health Organization country consultation.

\subsubsection{Abortion data classification}
In the past, abortion data were classified as incomplete based on expert assessment, whereas data were entered into our new model using a data classification algorithm. Our approach increases transparency. However, where there is insufficient evidence to classify a country's official statistics as incomplete, it is possible that our data classification process may cause our model to under-estimate the true incidence of abortions.

\paragraph{Germany and Japan}
The previous abortion study treated Germany's and Japan's data as incomplete, whereas we do not based on our data classification algorithm. We note these countries both because of this change and because they were flagged by the out-of-sample validation exercises. In Japan, although surgical abortions are available, mifepristone is not registered, and news reports discuss abortion under-reporting for reasons including stigma and tax evasion.\footnote{Goto A, Fujiyama-Koriyama C, Fukao A, Reich MR. Abortion trends in Japan, 1975-95. Studies in Family Planning. 2000 Dec; 31(4): 301-8.} In Germany, prior to re-unification, abortion was legal in East but not West Germany.\footnote{Eddy, Melissa. ``A Hitler-era abortion law haunts Merkel, and Germany.'' New York Times. March 27 2018.} Following re-unification, a law was passed legalizing abortion. The courts subsequently found this to conflict with the German constitution. However, they acknowledged the government's right to choose not to prosecute. German official statistics for the early 1990s states that abortions were under-enumerated, but subsequent reports describe them as complete.\footnote{Statistisches Bundesamt. Schwangerschaftsabbr\"{u}che, 1996-2018} 

\subsubsection{Intention data classification}

Our process for classifying intention data generally corresponds to the process used in Bearak et al, with two differences. If a survey sampled only ever-married women, the previous study treated the percent of births unintended as a non-representative datum. In this study, in contrast, we used survey micro-data to compute a range for the percent of marital births unintended, and we used the overall datum as the minimum percent unintended across all women. Also, for DHS and RHS reports where we did not have access to the micro-data, such that the inputs were for a 3- or 5-year recall periods, in contrast to the 1-year recall periods we used when we had access to the survey micro-data, our model included bias terms to adjust for this.




\section{Model validation}

\subsection{Excluding 20\% of all data classes at random}

Model performance was first assessed through out-of-sample validation exercises, whereby 20\% of the observations within each data class were excluded at random. For these exercises, data on official statistics were grouped by country-periods (sampling out all a country's official statistics within a five-year period), so that data from adjacent years within the same five-year period would not inform the  estimates produced from the training set.

To quantify model performance, we calculated various validation measures based on the excluded observations. The considered measures were based on prediction errors for the abortion rate and the percent of births unintended, where a prediction error refers to the difference between the excluded observation and the median of its posterior predictive distribution based on the training set, with error $e^y_i = 100 \cdot (p^y_i - \tilde{p}^y_i)$ for data on the proportion of births unintended and $e^a_j = 100 \cdot (p^a_j - \tilde{p}^a_j)$ for data on the abortion rate, where $\tilde{p}^y_i$ and $\tilde{p}^a_j$ denote the posterior median of the predictive distribution (based on the training set) for the excluded datum on the proportion of births unintended or on the abortion rate, respectively. Relative errors are given by $e_i/\tilde{p}_i$.

We also assessed the coverage of 80\% prediction intervals (to quantify the calibration of the prediction intervals). Coverage for birth intention is given by $1/N \sum 1[p^y_i \geq  l^y_i] \cdot 1[p^y_i \leq u^y_i]$, where $N^y$ denotes the total number of left-out observations considered and $l_i$ and $u_i$ the lower and upper bound of the 80\% prediction interval for the $i$-th excluded observation. Corresponding, for abortion rates, $1/N \sum 1[p^a_j \geq  l^a_j] \cdot 1[p^a_j \leq u^a_j]$, with $N^a$.

We compared the uncertainty bounds for estimates from the training set to the data-driven estimates (which are based on the full data set). The goal is to check that additional data does not change the current model-based estimates significantly; if more data becomes available, we expect model-based estimates produced from a larger dataset to lie well within the previously constructed uncertainty intervals. The smaller the proportion of estimates that fall outside their respective uncertainty intervals, the better.

We report results from these exercises in Table 6 (on the next page), for each of the reported indicators -- unintended pregnancy rates, the percent of unintended pregnancies ending in abortion, and abortion rates. For each of these indicators, the first rows, in bold, show the overall results. Below these, we also report the results separately by region.

For all observations combined, we find that the prediction intervals are calibrated as expected; 78.3\%--81.8\% of the excluded points lie within the prediction intervals (rows labeled ``Abortion data (rates per 1,000 women 15-49)'' and ``Intention data (\% unintended)'', in bold). Furthermore, when treating the data-driven estimates as the truth, we see that coverage increases to 85.2\% and 92.2\%, respectively, for abortion rates and the percentages of births unintended (rows labeled ``Abortion rates (corresp. to abortion points)'' and ``Percent of births unintended (corresp. to intention points)''). Finally, given that in our model, abortion and intention data inform both outcomes, we also took summarized results for country-years corresponding to abortion \textbf{or} intention data points. The results (rows ``Abortion rates (corresp. to abortion or intention points)'' and ``\% unintended (corresp. to abortion or intention points)'') also show the model to be well calibrated, with approximately 94\% coverage for both outcomes (93.5\% and 94.5\%, respectively). This is well above nominal (80\%). Finally, we found that the median errors across all countries were close to 0 in all exercises, ranging from -0.4 to 0.

When analyzing validations separately by region, results are more nuanced. As a reminder, it is preferred that no more than 10\% of the observations fall below, or above, the intervals. Based on predicting left out observations, we see that the lower bounds of the prediction intervals for the Central Europe, Eastern Europe, Central Asia and Caucasus (hereafter, ``Revised Eastern Europe'') region fall above the abortion points 15.6\% of the time, and that for the intention points the lower bounds of the prediction intervals lie above the observed points in the High-income, Revised Eastern Europe, South/east Asia and Oceania (hereafter, ``Asia''), and Middle East/North Africa (hereafter, ``MENA'') regions 21.6\%, 22.7\%, 14.3\%, and 25.4\% of the time, respectively. However, these percentages are based on a small average number of left out observations only. Moreover, when treating the data-driven estimates as the truth, we find the abortion UI's to contain the estimates at the desired coverage across all regions. In this exercise, we find that at most 8.7\% of the data-driven estimates fall below the intervals. Coverage, moreover, ranges from 82.9\% to 89.4\%. 

For the percentages of births unintended, validation by region based on left out observations suggest some possible bias in the High-income and Revised Eastern Europe regions when considering intention point inputs only, however errors on the proportion scale are small; moreover, validation results for this outcome based on  all point data show that the estimates are well-calibrated,  with better than nominal coverage of 95.5\% and 89.6\% respectively (rows below ``Abortion rates (corresp. to abortion or intention points), and, likewise, for intention data (rows below ``\% unintended (corresp. to abortion or intention points'').

The last series in the validation table takes into account that the estimated abortion rates are informed by intention data as well as abortion data, and likewise, the estimated percentages of births unintended are informed by abortion data as well as intention data, and the availability and quality of different data classes can vary systematically by region. In this exercise, we find the UIs to be well-calibrated in all world regions. Coverage for abortion ranged from 87.8\% (in the High-income region) to 100\% (in the MENA region). For intention, coverage ranged from 86.3\% (in the MENA region) to 97.7\% (in Sub-Saharan Africa). An exception to this general pattern was that in the MENA region, coverage was slightly less than nominal in that, while overall coverage was 86.3\%, such that no more than 13.7\% of the training-set UIs excluded the data-driven estimates, all of these were below, and none above, the intervals.

\includegraphics[scale=0.43,trim=4cm 2cm 0cm 2cm]{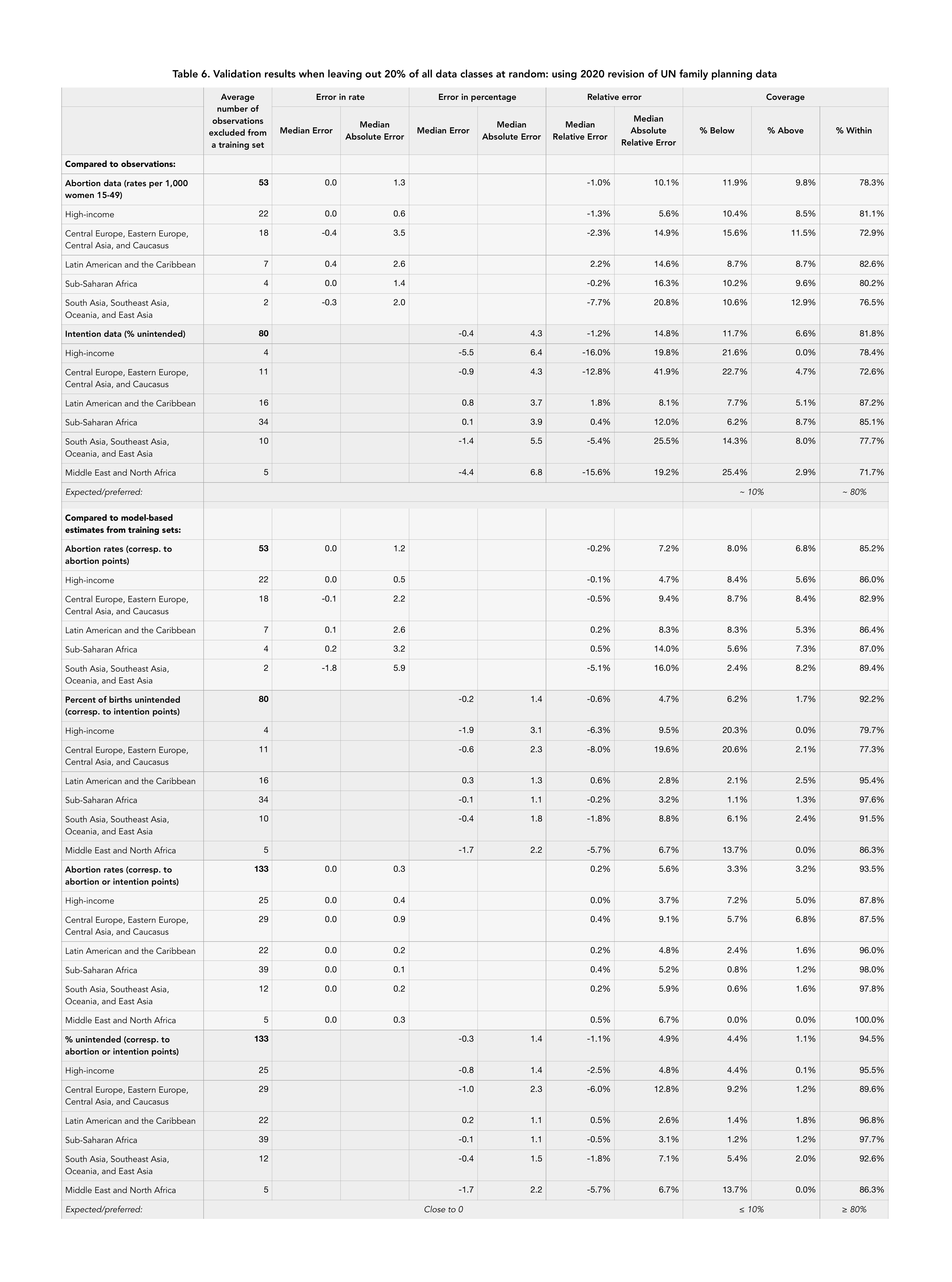}

\subsection{Leave-one-country-out validations}

To further quantify the performance of our final model, we also performed leave-one-country-out validations. In these exercises, we re-estimated the model once for each of the countries with data, and, in each of these runs, we excluded all data from a country. When calculating validation measures, we first averaged results (i.e. errors and coverage) within countries, to then average across countries, such that countries with multiple data points were not weighted more heavily than countries with one data point.

Validation results for the reported indicators -- unintended pregnancy rates, the percentage of unintended pregnancies that end in abortion, and abortion rates -- are presented in Table 7 (on the next page). For the unintended pregnancy rates and the percentage of unintended pregnancies that end in abortion, we report averages across all countries with abortion or intention point inputs, as well as reporting these validation results separately for world regions. For the abortion rates, we additionally report averages for only those countries with abortion point inputs, and for countries with point inputs for both outcomes.

Averaging across all countries with point inputs, we find that the uncertainty intervals are well calibrated for all indicators. Coverage is 90.4\% for the unintended pregnancy rates, 86.3\% for the percentage of unintended pregnancies that end in abortion, and 89.8\% for the abortion rates. Analyzing the subset of these countries that specifically have abortion point inputs, coverage is also beyond nominal, 82.2\%. Finally, for countries with point inputs for both outcomes, we see that coverage for the abortion rates is 88.9\%. We additionally find that median errors are close to zero for all outcomes. 

For the reported indicators, estimates appear to be unbiased, with well-calibrated uncertainty intervals, however relative differences across indicators are evident in the magnitudes of the residuals. The median absolute relative error (MARE) for the unintended pregnancy rates and the percentages of unintended pregnancies ending in abortion are close, at 11.5\% and 12.7\%, respectively. For the estimated abortion rates, the corresponding figure is also similar , at 13.5\%, although when averaging the absolute relative errors only over those countries with abortion point inputs, the MARE is 29\%.

Examining variation in these leave-one-country-out validation exercises separately by region,  the results are in general nominal or better, however we also note results that could be interpreted as evidence of over-estimation in the High-income region. There, averaging over countries with abortion point inputs, we see that the MRE is -12.3\%, and that 18.4\% of the data-driven country estimates fall below the UI's estimated from the training sets. In this region, however, we suspect that these results are driven by conservative decisions in the data classification process. Countries which were flagged by this exercise include Japan and Germany, whose abortion statistics had been classified as incomplete in previous studies (see: \nameref{sec:diffs}). We also note that the median relative error for Latin American countries with abortion point inputs is 33.9\%, which is suggestive of conservative estimates. However the data-driven estimates lie above the UI's only 1.2\% of the time, and across Latin American countries with abortion or intention point inputs, median relative error is only -1.5\%.

\includegraphics[scale=0.40,trim=6cm 0cm 0cm 0cm]{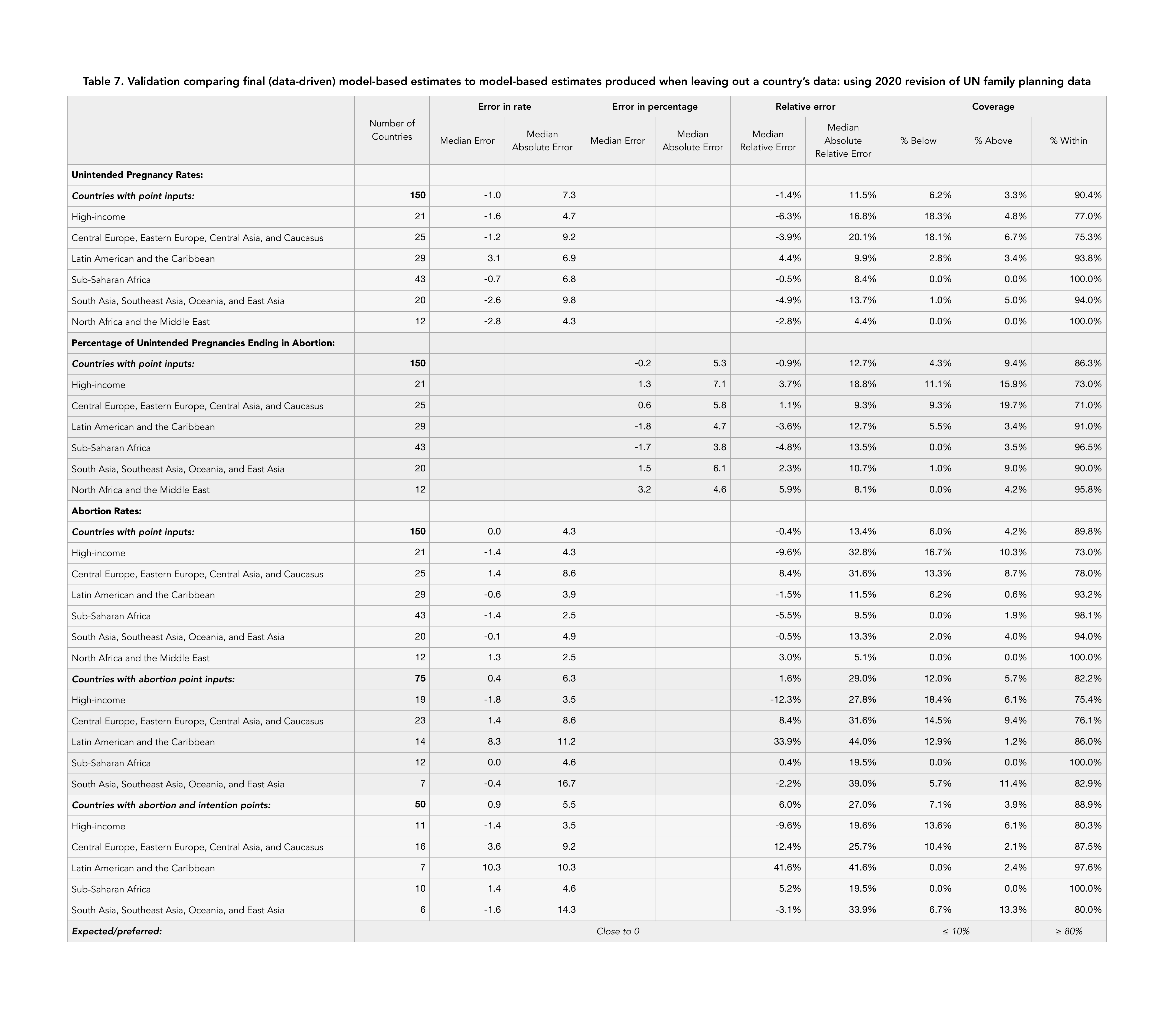}
\clearpage

\section{Appendix}
\setcounter{table}{0}
\begin{table}
    \includegraphics[scale=0.33,trim=4cm 0cm 0cm 0cm]{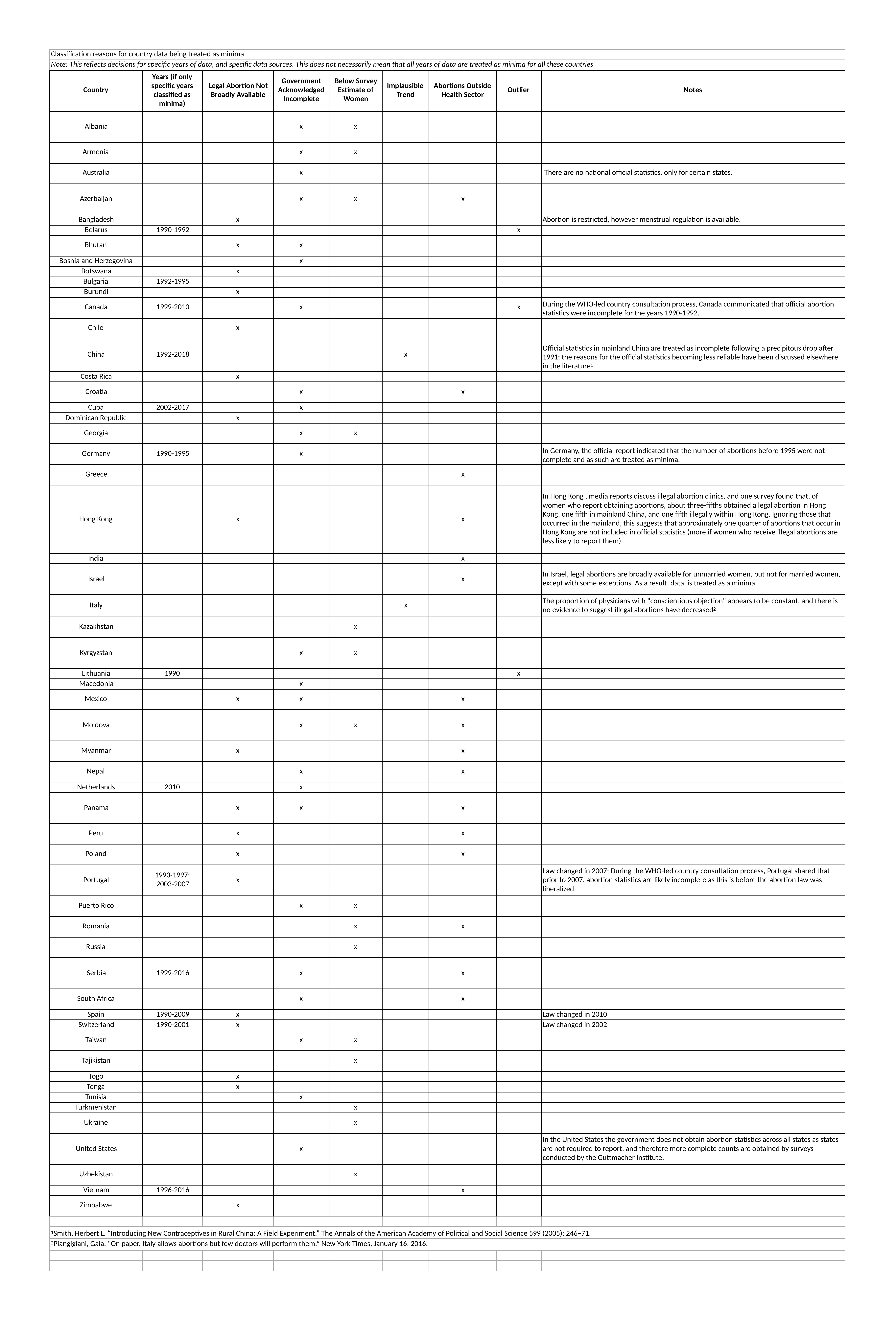}
    \caption{Classification reasons for country data treated as minima.}
    \label{tb:minima}
\end{table}

\end{document}